\documentclass[12pt]{article}
\usepackage{amsthm,amsfonts,amssymb,latexsym,amsmath}

{\catcode `\@=11 \global\let\AddToReset=\@addtoreset}
\AddToReset{equation}{section}

\theoremstyle{plain}
\newtheorem{thm}{THEOREM}
\newtheorem{cor}{COROLLARY}
\newtheorem{lem}{LEMMA}
\newtheorem{prop}{PROPOSITION}
\theoremstyle{definition}
\newtheorem{rem}{REMARK}
\newtheorem{defn}{DEFINITION}

\newcommand{\infspec}{{\rm inf\ spec\ }}
\newcommand{\R}{{\mathbb R}}
\newcommand{\C}{{\mathbb C}}
\newcommand{\N}{{\mathbb N}}

\newcommand{\Q}{{\mathcal Q}}
\newcommand{\HH}{{\mathbf H}}
\newcommand{\eps}{\varepsilon}
\newcommand{\B}{{\cal B}}

\newcommand{\x}{{\bf x}}

\newcommand{\Tr}{{\rm Tr}}
\newcommand{\half}{\mbox{$\frac{1}{2}$}}
\newcommand{\E}{{\mathcal E}^{\rm GP}}
\newcommand{\En}{{\mathcal E}_n}
\newcommand{\Engp}{E^{\rm GP}}
\newcommand{\Endm}{E^{\rm DM}}
\newcommand{\Edm}{{\mathcal E}^{\rm DM}}
\newcommand{\rdm}{\rho^{\rm DM}}
\newcommand{\gdm}{\gamma^{\rm DM}}
\newcommand{\mdm}{\mu^{\rm DM}}
\newcommand{\Hdm}{H^{\rm DM}}
\newcommand{\nc}{{n_c}}
\newcommand{\ndm}{{n^{\rm DM}}}

\begin{document}

\markboth{}{\scriptsize{R. SEIRINGER,
ROTATING BOSE GAS}}
\title{\bf{Gross-Pitaevskii Theory of the Rotating Bose Gas}}
\author{\vspace{5pt} Robert Seiringer\thanks{Erwin Schr\"odinger Fellow. On leave
from Institut f\"ur Theoretische Physik, Universit\"at Wien,
Boltzmann\-gasse 5, 1090 Vienna, Austria}\\
\vspace{-4pt}\small{Department of Physics, Jadwin Hall, Princeton
University}\\ \vspace{-4pt} \small{P.O. Box 708, Princeton, New
Jersey 08544, USA}\\
\small{{\texttt{rseiring@math.princeton.edu}}}}
\date{\small April 17, 2002}
\maketitle

\begin{abstract}
We study the Gross-Pitaevskii functional for a rotating
two-dimen\-sional Bose gas in a trap. We prove that there is a
breaking of the rotational symmetry in the ground state; more
precisely, for any value of the angular velocity and for large
enough values of the interaction strength, the ground state of the
functional is not an eigenfunction of the angular momentum. This
has interesting consequences on the Bose gas with spin; in
particular, the ground state energy depends non-trivially on the
number of spin components, and the different components do not
have the same wave function. For the special case of a harmonic
trap potential, we give explicit upper and lower bounds on the
critical coupling constant for symmetry breaking.
\end{abstract}

\section{Introduction}

We consider the Gross-Pitaevskii (GP) theory of a rotating
two-dimensional Bose gas in a trap. The Bose gas is described by a
single function $\phi$ on $\R^2$, the wave function of the
condensate. It is confined in some trap potential $V$, and rotates
around the origin at an angular velocity $\Omega$. The strength of
the interaction between the particles is measured by the positive
parameter $a$ appearing in the GP functional (\ref{gpfunct})
below. It is related to the particle number $N$, the scattering
length $a_s$ of the interaction potential and the average particle
density $\rho$ via
\begin{equation}
a=\frac {4\pi N}{|\ln a_s^2\rho|}.
\end{equation}
Minimization of the GP functional is supposed to describe the physical
properties of rotating Bose gases at very low temperatures, as
considered in recent experiments \cite{MCWD00,ARVK01}. These show
various interesting properties, in particular, the appearance of
multiple vortices and a resulting breaking of the rotational
symmetry. There have been a lot of theoretical investigations on these
phenomena (see, e.g., the review article \cite{FS01}), based on the GP
approach, either using numerical methods or various simplifying
approximations, but a proof that the Gross-Pitaevskii functional
indeed captures all these features is still missing.

We shall not be concerned here with the derivation of the GP
functional from the basic quantum mechanical $N$-particle
Hamiltonian. For non-rotating systems, i.e., $\Omega=0$, this has been
achieved in \cite{LSY00,LSY01} (see also \cite{LS02}). However, the
methods used there allow no simple generalization of these results to
the rotating case.

We shall now describe the setting more precisely. We denote by
$(r,\varphi)$ polar coordinates for $\x=(x,y)\in \R^2$. In these
coordinates, the angular momentum is given by $L=-i
\partial/\partial\varphi$.  For
\begin{equation}
H_0=-\Delta-\Omega L + V(r)
\end{equation}
and $\phi\in \Q(H_0)\cap L^4(\R^2,d^2\x)$ define the {\em
Gross-Pitaevskii energy functional} by
\begin{equation}\label{gpfunct}
\E[\phi]=\langle \phi| H_0 \phi\rangle + a \int|\phi(\x)|^4 d^2\x.
\end{equation}
(Here $\Q(H_0)$ denotes the quadratic form domain of $H_0$, and
$\langle\cdot|\cdot\rangle$ denotes the standard inner product on
$L^2(\R^2)$). The parameter $a$ is non-negative, and without loss
of generality also $\Omega\geq 0$. We assume that $V\in
L^\infty_{\rm loc}(\R^2)$ is a positive radial function with the
property that 
\begin{equation}\label{propv}
V(r)\geq \tilde\Omega^2 r^2/4-C_{\tilde\Omega}
\end{equation}
for all $\tilde\Omega$ in some non-zero interval $[0,\Omega_c)$, with
$C_{\tilde\Omega}<\infty$. We take $[0,\Omega_c)$ to be the largest
such interval, allowing it to be the whole half line; i.e., we allow
$\Omega_c$ to be infinity. Moreover, we assume that $V(r)$ is
polynomially bounded at infinity, i.e, there exist constants $C_1$,
$C_2$ and $2\leq s<\infty$ such that $V(r)\leq C_1+ C_2 r^s$. For
convenience, let $\inf_r V(r)=0$.

Let $\Engp(a,\Omega)$ be the ground state energy of $\E$, i.e,
\begin{equation}
\Engp(a,\Omega)=\inf\left\{\E[\phi], \phi\in\Q(H_0)\cap L^4(\R^2),
\|\phi\|_2=1\right\},
\end{equation}
being finite for $|\Omega|< \Omega_c$ and $a\geq 0$. Using
standard methods (see e.g. \cite{LSY00}) one can show that there
exists a minimizer $\phi^{\rm GP}$ for $\E$ as long as
$|\Omega|<\Omega_c$, i.e., the infimum is actually a minimum. For
$|\Omega|>\Omega_c$ the functional $\E$ is not bounded from below.

The purpose of this paper is a detailed study of the GP functional
(\ref{gpfunct}). One of our main results is that for any $\Omega>0$
and for large enough interaction strength $a$, {\em no} minimizer of
$\E$ is an eigenfunction of the angular momentum $L$. Since $\E$ is
invariant under rotation of $\phi$, this result means that the
rotational symmetry is broken in the ground state. This has
interesting consequences on the multi-component Bose gas (or
equivalently, Bose gas with spin). In particular, we will show that
the ground state energy (of the natural generalization of the GP
functional to multi-component systems) depends non-trivially on the
number of spin components, and the different components necessarily
have a different wave function in the symmetry breaking regime.

The paper is organized as follows: In Section \ref{sectvort} we
study stationary points of the GP functional, in particular
minimizers of $\E$ restricted to the subspace of eigenfunctions of
the angular momentum with fixed eigenvalue $n$, so-called vortex
states. We show that for large enough angular momentum, these
vortex states can never be the absolute minimizer of the GP
functional, uniformly in the coupling constant $a$, which will be
crucial in the proof of symmetry breaking. In Section
\ref{sectcrit} we study the critical values of the angular
velocity $\Omega$ for that an $n+1$-vortex becomes energetically
favorable to an $n$-vortex. These critical velocities all tend to
zero as $a$ goes to infinity, which will allow us to conclude that
all vortex states with angular momentum smaller than a certain
value cannot be the actual minimizers of $\E$. In Section
\ref{sectsymm} we will use these results to prove symmetry
breaking. Section \ref{sectdm} is devoted to the study of a GP
density matrix functional, which will be useful in investigations
on the multi-component Bose gas in Section \ref{sectmult}. There
we show that in the symmetry breaking regime, the GP energy
depends non-trivially on the number of spin components. In Section
\ref{sectharm} we finally consider the special case of an harmonic
potential $V(r)=r^2$, where we derive explicit upper and lower
bounds on the critical coupling constant for symmetry breaking.

\section{Vortex states}\label{sectvort}

Given any stationary state of $\E$, i.e., a function $\phi$ with
$\|\phi\|_2=1$ satisfying
\begin{equation}\label{station}
\left( H_0 + 2a |\phi|^2 - \mu\right)\phi = 0
\end{equation}
for some $\mu\in\R$, we define the (real) quadratic form $Q(w)$ by
the perturbation
\begin{equation}\label{perturb}
\E_\mu[\phi+\eps w]-\E_\mu[\phi]= \eps^2 Q(w) + O(\eps^3)
\end{equation}
as $\eps\to 0$, where $\E_\mu[\phi]=\E[\phi]-\mu \int |\phi|^2$. A
simple calculation shows that
\begin{equation}
Q(w)=\langle w| H_0 + 4 a |\phi|^2 - \mu |w\rangle + 2 a\, \Re
\int\overline{\phi ^2} w^2,
\end{equation}
where $\Re$ denotes the real part. 
Multiplying (\ref{station}) with $\phi$ and integrating shows that
\begin{equation}
\mu=\E[\phi]+a\int|\phi|^4.
\end{equation}

\begin{defn}[Stability]
We say that a stationary state $\phi$ is stable if and only if
$Q(w)\geq 0$ for all $w\in \Q(H_0)\cap L^4(\R^2)$ that are orthogonal
to $\phi$.
\end{defn}

We have $Q(i\phi)=0$, which corresponds to a simple phase change
in (\ref{perturb}). Moreover, $Q(\partial\phi/\partial\varphi)=0$
because of rotational invariance of $\E$. Note that, by
definition, an absolute minimizer of $\E$ is necessarily stable, since
\begin{equation}
\E[(\phi+\eps w)/\|\phi+\eps w\|_2]= \E[\phi] + \eps^2 Q(w) + O(\eps^3)
\end{equation}
if $\langle\phi|w\rangle=0$.

We now look for special solutions to (\ref{station}) of the form
\begin{equation}\label{vortex}
\phi(\x)=f(r)e^{i n \varphi}
\end{equation}
for some $n\in \N$, a so-called {\it n-vortex}. Here $f$ is a real
radial function. Since
$\E[fe^{-in\varphi}]=\E[fe^{in\phi}]+2n\Omega$ we can restrict
ourselves to non-negative $n$ without loss of generality. At least
one solution of the form (\ref{vortex}) for each $n$ always
exists, as one easily sees by minimizing the functional $\E$ in
the subspace of functions with $L\phi=n\phi$. For $\phi$ of the
form (\ref{vortex}) there is the following direct sum
decomposition of $Q$: Writing $w(\x)=\sum_{m\geq 0} w_m(\x)$ with
\begin{equation}
w_m(\x)= A_m(r) e^{i (n-m)\varphi}+B_m(r) e^{i (n+m)\varphi}
\end{equation}
one easily sees that $Q(w)=\sum_{m\geq 0} Q(w_m)$.

Considering the stability of vortices, we restrict ourselves to
the case $\Omega<\Omega_c$, since for $\Omega> \Omega_c$ all
states are certainly unstable. (The case $\Omega=\Omega_c$ depends
on the particular form of the potential $V$.) First of all, $\phi$
has only a chance of being stable if $f$ has no zeros away from
$r=0$. More precisely, the following proposition holds.

\begin{prop}[Instability for $f$'s with zeros]
Assume that either $n=0$ and $f$ has some zero, or $n\geq 1$ and
$f$ has some zero away from $r=0$. Then $\phi$ is unstable.
\end{prop}

\begin{proof}
For some real $h$ we choose $w(\x)=i h(r)e^{in\varphi}$ as a trial
function for $Q$. We get
\begin{equation}
Q(w)=\langle
h|-\Delta+\frac{n^2}{r^2}-n\Omega+V(r)+2af^2-\mu|h\rangle\equiv
\langle h|\widetilde H|h\rangle.
\end{equation}
We know that $\widetilde H f=0$, but because of its zeros, $f$
cannot be the ground state of $\widetilde H$, so there exists an
$h$, orthogonal to $f$, with $\langle h|\widetilde H|h\rangle<0$.
\end{proof}

Now let $\phi$ be the minimizer of $\E$ in the subspace with
angular momentum $n$. Then $f$ defined in (\ref{vortex}) minimizes
the energy functional
\begin{equation}\label{defen}
\En[f]=\langle f|-\Delta + \frac {n^2}{r^2}+V(r)|f\rangle + 2\pi a
\int_0^\infty |f(r)|^4 r dr
\end{equation}
under the condition $2 \pi \int |f(r)|^2 r dr =1$, with
corresponding energy
\begin{equation}\label{26}
E_n(a)=\En[f]=\E[f e^{i n\varphi}]+\Omega n.
\end{equation}
In the following, we will study $\En$ for all $n\geq 0$, not only
for integers. Denote
\begin{equation}
\tilde\mu\equiv \mu+n\Omega,
\end{equation}
which is independent of $\Omega$. The minimizer $f$ of $\En$ has
the following properties.

\begin{lem}[Properties of $f$]\label{proplem}
$f(r)>0$ for $r>0$, $f\in C^\infty(\R_+)$ if $V\in C^\infty$, and
$f(r)=O(r^n)$ as $r\to 0$. Moreover, $f\in L^\infty(\R_+)$, and
\begin{equation}\label{fsup}
\|f\|_\infty^2\leq \frac {\tilde\mu} {2a}.
\end{equation}
\end{lem}

\begin{proof}
The regularity and strict positivity follow in a standard way from the
variational equation for $f$ (cf. \cite{LL01}). Writing $f(r)=r^n
g(r)$ we see that $g$ minimizes the functional
\begin{equation}\label{gfunc}
\tilde {\mathcal E}[g]=\int_0^\infty
g(r)\left(-g''(r)-\frac{2n+1}{r}g'(r)+V(r) g(r)+a r^{2n}
g(r)^3\right) r^{2n+1} dr
\end{equation}
under the condition $2\pi \int g(r)^2
r^{2n+1}dr=1$, from which we conclude that $g$ is a bounded,
strictly positive function.

The bound (\ref{fsup}) is proved analogously to Lemma 2.1 in
\cite{LSY01}: Let $\B=\{\x, 2 a f(|\x|)^2> \tilde\mu\}$. We see
that $-\Delta f<0 $ on $\B$, i.e., $f$ is subharmonic on $\B$ and
therefore achieves its maximum on the boundary of $\B$. Hence $\B$
is empty.
\end{proof}

We remark that all the properties of $f$ stated in Lemma
\ref{proplem}, except for the positivity, hold for all
$n$-vortices and not only for minimizers. Also the following lemma
holds true for arbitrary vortex states.

\begin{lem}[Properties of $g$]\label{propg}
Let $f(r)=r^n g(r)$ be a stationary point of $\En$, for $n\in
\R_+$. Then
\begin{equation}\label{gbound}
\|g\|_\infty\leq \|f\|_\infty \left(c_n^2 \tilde\mu\right)^{n/2},
\end{equation}
where
\begin{eqnarray} \nonumber
&&c_n=\left(\frac{2^{-n}\left(\frac{2-n}n\right)^{n/2}\pi\,{\rm
Csc}\left(\frac{n\pi}2\right)} {(2-n)\Gamma(n)} \right)^{1/n}
\quad\mbox{for $n\leq 1$} \\ \label{cn}
&&c_n=\frac{\sqrt\pi}n\frac{\Gamma(n+\half)} {\Gamma(n)}
\quad\mbox{for $n\geq 1$.}
\end{eqnarray}
If $f$ is the minimizer of $\En$, and if $V$ is monotone
increasing, then $g$ is a monotone decreasing function.
\end{lem}

\begin{proof}
By a rearrangement argument one sees from (\ref{gfunc}) that the
minimizer of $\tilde {\mathcal E}$ is monotone decreasing if $V$
is monotone increasing. For a general $n$-vortex, $g$ fulfills the
equation
\begin{equation}
-g''(r)-\frac {2n+1}r g'(r)+V(r)g(r)+2a r^{2n} g(r)^3=\tilde\mu
g(r).
\end{equation}
Kato's inequality and the positivity of $V$ imply that
\begin{equation}\label{varg}
-|g(r)|''-\frac{2n+1} r |g(r)|'\leq \tilde\mu |g(r)|
\end{equation}
in the sense of distributions. Now let $\chi_n(r,s)$ be the kernel
of the operator
\begin{equation}\label{defop}
\left(-\frac {d^2}{dr^2}-\frac{2n+1}r \frac d{dr}+1\right)^{-1},
\end{equation}
acting on $L^2(\R_+,r^{2n+1}dr)$. It is given by
\begin{equation}
\chi_n(r,s)=\frac 1{(rs)^n}\begin{cases} I_n(r) K_n(s) & \mbox{for
$r\leq s$}\\ K_n(r) I_n(s) & \mbox{for $r\geq s$,}
\end{cases}
\end{equation}
where $I_n$ and $K_n$ denote the usual modified Bessel functions.
Note that both $I_n$ and $K_n$ are positive, so $\chi_n$ is
positive. By scaling, the integral kernel of (\ref{defop}) with
$+1$ replaced by $+t^2$ is $t^{2n}\chi_n(rt,st)$. Therefore
(\ref{varg}) implies that, for $t>0$ and $0\leq\alpha\leq 1$,
\begin{eqnarray}\nonumber
|g(r)|&\leq& (\tilde\mu+t^2) t^{2n} \int_0^\infty \chi_n(rt,st)
|g(s)| s^{2n+1} ds\\ \label{green} &\leq& (\tilde\mu+t^2) t^{2n}
\|g\|_\infty^{1-\alpha}\|f\|_\infty^\alpha  \int_0^\infty
\chi_n(rt,st)s^{2n+1-n\alpha} ds.
\end{eqnarray}
We now claim that $\int\chi_n(r,s)h(s)s^{2n+1}ds$ is monotone
decreasing in $r$ if $h$ is a positive, monotone decreasing
function. To prove this, it suffices to consider a step function
$h(s)=\Theta(R-s)$, $R>0$. A simple calculation yields
\begin{equation}
\int_0^R \chi_n(r,s)s^{2n+1} ds=\begin{cases}
1-R^{n+1}K_{n+1}(R)\frac{I_n(r)}{r^n} & \mbox{for $r\leq R$} \\
R^{n+1}I_{n+1}(R)\frac{K_n(r)}{r^n} & \mbox{for $r\geq R$},
\end{cases}
\end{equation}
which proves the claim, since $I_n(r)/r^n$ and $K_n(r)/r^n$ are
monotone increasing and decreasing, respectively. Therefore the
maximum on the right hand side of (\ref{green}) is achieved for
$r=0$, and we get
\begin{equation}
\|g\|_\infty\leq \frac{\tilde\mu+t^2}{t^{2-n\alpha}}
\|g\|_\infty^{1-\alpha} \|f\|_\infty^\alpha \frac {2^{-n}}{
\Gamma(n+1)}\int_0^\infty K_n(s) s^{n+1-n\alpha} ds.
\end{equation}
The last integral can be evaluated explicitly, if $n\alpha <2$.
Choosing $\alpha=1$ for $n\leq 1$ and $\alpha=1/n$ for $n\geq 1$
and optimizing over $t$ yields the desired result.
\end{proof}

Equation (\ref{gbound}) effectively gives a lower bound on $s$,
the size of the vortex core, defined by $|f(r)|\sim \|f\|_\infty
(r/s)^n$ as $r\to 0$, i.e.,
\begin{equation}
s=\left(\lim_{r\to 0} \frac {|f(r)|}{r^n
\|f\|_\infty}\right)^{-1/n}\geq\left(
\frac{\|f\|_\infty}{\|g\|_\infty}\right)^{1/n}\geq\frac
1{\tilde\mu^{1/2} c_n}.
\end{equation}
Note that $c_n^n\to 1$ as $n\to 0$, and $c_n=O(n^{-1/2})$ for
large $n$. The latter fact will be important in the proof of the
following theorem.

\begin{thm}[Instability for large $n$]\label{largenthm}
For all $0\leq \Omega<\Omega_c$ there exists an $N_\Omega<\infty$
(independent of $a$!) such that all vortices with $n\geq N_\Omega$
are unstable.
\end{thm}

\begin{proof}
Let $n\geq 1$, and let $w_1\in H^1(\R^2)$ be radial and
normalized, with support in the ball of radius $1$. Let $X=\int
|w_1(r)|^2 V(r/c_n\tilde\mu^{1/2})d^2\x$, $T=\langle w_1|-\Delta
w_1\rangle$, and define $w$ by
\begin{equation}
w(\x)=c_n \tilde\mu^{1/2} w_1(c_n\tilde\mu^{1/2} r),
\end{equation}
with
$c_n$ given in (\ref{cn}). We have, using (\ref{fsup}) and
(\ref{gbound}),
\begin{equation}
Q(w)\leq n\Omega+\tilde\mu\left(c_n^2 T+\frac
1{\tilde\mu}X-1+2\int |w(\x)|^2 \min\{1,(r^2\tilde\mu c_n^2)^n\}
d^2\x\right).
\end{equation}
With $M=\int |w_1|^2r^{2n}$ this gives
\begin{equation}\label{qtnm}
Q(w)\leq n\Omega+\tilde\mu\left(c_n^2 T-1+2M\right)+X.
\end{equation}
Now $\tilde\mu$ is larger than $e_n^{(0)}\equiv\infspec
(-\Delta+V)\restriction_{L=n}$, which can be bounded below as
follows. By assumption (\ref{propv}), $V(r)\geq
-C_{\tilde\Omega}+\tilde\Omega^2 r^2/4$ for some constant
$C_{\tilde\Omega}$ and $\Omega<\tilde\Omega<\Omega_c$. Denoting by
$\psi_n^{(0)}$ the eigenfunction corresponding to $e_n^{(0)}$, we
have
\begin{equation}
e_n^{(0)}\geq \langle\psi_n^{(0)}|\frac
{n^2}{r^2}+V(r)|\psi_n^{(0)}\rangle\geq
\langle\psi_n^{(0)}|-C_{\tilde\Omega}+\frac
{n^2}{r^2}+\frac{\tilde\Omega^2}4 r^2 |\psi_n^{(0)}\rangle \geq
-C_{\tilde\Omega}+n\tilde\Omega.
\end{equation}
Now $c_n^2=O(n^{-1})$ as $n\to\infty$, and the same holds for $M$.
$X$ can be bounded by $\sup_{r\leq (c_n^2\tilde\mu)^{-1/2}}V(r)$,
which is, for fixed $\Omega$, bounded independent of $n$ and $a$
by the considerations above. Therefore $Q(w)<0$ for $n$ large
enough.
\end{proof}

For a special class of potentials, we can extend the previous
result in the following way:

\begin{thm}[Instability for special $V$'s]\label{instthm}
Assume that for some $d\in\N$, $d\geq 2$,
\begin{equation}\label{condv}
\left(r\left(\frac {V(r)}{r^{2(d-1)}}\right)'\right)'\leq 0
\end{equation}
for all $r$. Assume also that $n\geq d$ and
\begin{equation}\label{condmu}
\tilde\mu>n\Omega\left(1+\frac 2{d-1}\right).
\end{equation}
Then $\phi$ is unstable.
\end{thm}

\begin{proof}
For $1\leq d\leq n$ we choose as a trial function
\begin{equation}
w(\x)=\left(A(r)+B(r)\right)e^{i(n-d)\varphi}
+\left(A(r)-B(r)\right)e^{i(n+d)\varphi}.
\end{equation}
Then $Q(w)$ can be written as
\begin{equation}
Q(w)=2\left\langle\begin{array}{c}A\\B\end{array}\right|\, \HH \,
\left|\begin{array}{c}A\\B\end{array}\right\rangle,
\end{equation}
where
\begin{equation}
\HH=\left(\begin{array}{cc} H_0+\frac{n^2+d^2}{r^2}-n\Omega-\mu+
6a f^2 & d\Omega - \frac{2nd}{r^2}
\\ d\Omega - \frac{2nd}{r^2} & H_0+\frac{n^2+d^2}{r^2}-n\Omega-\mu+ 2a f^2 \end{array}
\right).
\end{equation}
We now choose $A(r)=f'(r)/r^{d-1}$ and $B(r)=n f(r)/r^d$. Note
that $A-B=r^{n-d+1}(f/r^n)'\leq O(r)$ as $r\to 0$, so $w\in
H^1(\R^2)$. A straightforward calculation using Equation
(\ref{station}) yields
\begin{eqnarray}\nonumber
Q(w)&=&8\pi  \int_0^\infty dr \frac {f'(r)f(r)}{r^{2(d-1)}}
\left(-\mu(d-1) + 2a(d-1) f(r)^2 + n\Omega\right. \\ \nonumber &&
\hspace{50pt} \left.+(d-1)V(r)-\frac r2 V'(r) \right)
\\ \nonumber &=&8 \pi \int_0^\infty dr \frac {f(r)^2}{r^{2(d-1)+1}} \left(-\mu (d-1)^2+
a(d-1)^2 f(r)^2 + (d-1)n\Omega\right. \\ &&
\hspace{50pt}\left.+\frac 14 r^{2(d-1)+1}\left(r\left(\frac
{V(r)}{r^{2(d-1)}}\right)'\right)' \right), \label{227}
\end{eqnarray}
where we used partial integration in the last step. Estimating $a
f(r)^2$ by (\ref{fsup}) this shows the negativity of $Q(w)$ as
long as (\ref{condv}) and (\ref{condmu}) are satisfied.
\end{proof}

In the case of a homogeneous potential $V(r)=r^\nu$, $2\leq
\nu<\infty$, the condition (\ref{condv}) is fulfilled for
$\nu=2(d-1)$, $d\in\N$, showing that in this case every vortex with
$n\geq d=\half\nu+1$ is unstable, if (\ref{condmu}) is fulfilled,
i.e., if $\tilde\mu$ is large enough. (For fixed $n$ this means that
$a$ has to be large enough.)

\begin{rem}[Translational stability]\label{trans}
The calculation in the proof of Theorem \ref{instthm} shows that
any vortex with $n\geq 1$ is stable against translations, if the
opposite of the assumption on $V$ is true for $d=1$. More
precisely, the function $w(\x)$ defined above is, for $d=1$, equal
to $\partial \phi/\partial x$. Looking at (\ref{227}) we see that
this expression is always positive for $d=1$, if $(r V'(r))'\geq
0$. This implies that $Q(\partial \phi/\partial x)\geq 0$, and the
same conclusion holds for $\partial \phi/\partial y$. Note that
the condition on $V$ is in particular fulfilled for any
homogeneous potential $V(r)=r^\nu$.
\end{rem}

The choice of the test function in the proof of Thm. \ref{instthm}
is motivated by analogous considerations in \cite{H82} (see also
\cite{M95} for a treatment of the Ginzburg-Landau model).

\section{The critical frequencies}\label{sectcrit}

{}From (\ref{26}) one sees that an $n+1$-vortex becomes
energetically favorable to an $n$-vortex if $\Omega>\Omega_n$,
where the critical frequency is given by
\begin{equation}\label{critdef}
\Omega_n(a)=E_{n+1}(a)-E_n(a)>0,
\end{equation}
with $E_n(a)$ defined in (\ref{26}). In the following we will
study the properties of the $\Omega_n$'s, in particular their
behavior for large $a$. This will be important in the proof of
symmetry breaking in the ground state of the GP functional.

\begin{lem}[Relation between $\Omega_n$'s]\label{omegan0}
For $n\geq 0$
\begin{equation}
\Omega_{n+1}\leq \frac{2n+3}{2n+1} \Omega_n.
\end{equation}
\end{lem}

\begin{proof}
Using $f_{n+1}$, the minimizer for ${\mathcal E}_{n+1}$, as a trial function
for ${\mathcal E}_{n+2}$ and $\En$, respectively, we get
\begin{equation}
\Omega_{n+1}\leq (2n+3) \int \frac{f_{n+1}(r)^2}{r^2} d^2\x
\end{equation}
and
\begin{equation}\label{37}
\Omega_n \geq (2n+1) \int \frac{f_{n+1}(r)^2}{r^2} d^2\x.
\end{equation}
\end{proof}

\begin{thm}[Bounds on $\Omega_n$]\label{decrthm}
For all $n\in\N_0$
\begin{eqnarray} \label{omna}
&&\Omega_n(a)\leq (2n+1) \frac{2\pi e}a E_1(a)
\left(3+\left[\ln\left(\frac a{2\pi e^2}\right) \right]_+\right),
\\ &&\Omega_n(a)\geq (2n+1)\frac{1}4\frac
{\tilde\Omega^2}{C_{\tilde\Omega}+E_{n+1}(a)} \quad \mbox{ for all
$0\leq\tilde\Omega<\Omega_c$.}
\end{eqnarray}
\end{thm}

\begin{proof}
The concavity of $E_n(a)$ in $n^2$ implies that the right and left
derivatives of $E_n$ with respect to $n$ exist, and from the
existence of a unique minimizer for $\En$ for all $n$ we conclude
that $E_n$ is in fact differentiable in $n$. Therefore we have
\begin{equation}\label{omegade}
\Omega_n=\left. \frac{\partial E_n}{\partial n}\right |_{n=n_0}
\end{equation}
for some $n_0\in (n,n+1)$. Now let $f(r)=r^n g(r)$ be the
minimizer of $\En$. To obtain the upper bound, we estimate, for
$0<\alpha\leq 1$,
\begin{eqnarray}\nonumber \frac{\partial E_n}{\partial n}&=& 2n\int
\frac{f(r)^2}{r^2} d^2\x\\ \nonumber &\leq& 2n\left( \|f\|_4^2
\left(\int_{|\x|\geq R} r^{-4}
d^2\x\right)^{1/2}+\|g\|_\infty^{2\alpha}\|f\|_\infty^{2(1-\alpha)}
\int_{|\x|\leq R} r^{2n\alpha-2} d^2\x\right) \\ &\leq& 2n
\left(\|f\|_\infty \frac {\sqrt\pi}R +
\|g\|_\infty^{2\alpha}\|f\|_\infty^{2(1-\alpha)} \frac \pi
{n\alpha} R^{2n\alpha}\right) \end{eqnarray} for all $R>0$.
Optimizing over $R$ yields
\begin{equation}\label{parten} \frac{\partial E_n}{\partial n}\leq
\left(2n+\frac
1\alpha\right)\left(2\pi^{1+n\alpha}\|f\|_\infty^{2n\alpha+2(1-\alpha)}
\|g\|_\infty^{2\alpha} \right)^{1/(2n\alpha+1)},
\end{equation}
and by (\ref{gbound}),
\begin{equation}\label{asyom}
\frac{\partial
E_n}{\partial n}\leq \left(2n+\frac
1\alpha\right)\left(2\pi^{1+n\alpha}c_n^{2n\alpha}\|f\|_\infty^{2n\alpha+2}
\tilde\mu^{\alpha n} \right)^{1/(2n\alpha+1)}.
\end{equation}
Next we choose
\begin{equation}
\alpha=\min\left\{1,\frac 1n \left[\ln\left(\frac{c_n^2
\tilde\mu}{4\pi e^2 \|f\|_\infty^2}\right)\right]_+^{-1}\right\}
\end{equation}
and use (\ref{fsup}), which yields
\begin{equation}\label{312}
\frac{\partial E_n}{\partial n}\leq \frac{\pi e}a \tilde\mu
\max\left\{2n+1,n\ln\left(\frac{c_n^2 a}{2 \pi}\right)\right\}.
\end{equation}
Using $\tilde\mu\leq 2 E_n(a)$ and $c_n^{2n}\leq e$ this gives,
together with (\ref{omegade}), for $\Omega_0$
\begin{equation}
\Omega_0(a)\leq \frac{2\pi e}a E_1(a)
\max\left\{3,1+\ln\left(\frac a{2\pi}\right) \right\}.
\end{equation}
Now $\Omega_n\leq (2n+1)\Omega_0$ by Lemma \ref{omegan0}, which
finishes the proof of the upper bound.

To obtain the lower bound, we use (\ref{37}) and
\begin{equation}\label{314}
\int \frac{f_{n+1}^2}{r^2} d^2\x\geq \frac{1}{\int f_{n+1}^2 r^2
d^2\x}\geq \frac{\tilde\Omega^2}4\frac
1{C_{\tilde\Omega}+E_{n+1}(a)}
\end{equation}
because of (\ref{propv}), for all $0\leq\tilde\Omega<\Omega_c$.
\end{proof}

Note that since $V(r)\leq C_1+C_2 r^s$ for some $2\leq s<\infty$
by assumption, a simple trial wave function shows that $E_n(a)\leq
O(a^{s/(s+2)})$ as $a\to\infty$, implying that $\Omega_n$ behaves
at most as $a^{-2/(s+2)}\ln a$ for large $a$. In particular,
$\lim_{a\to\infty}\Omega_n(a)=0$ for all $n$.

\section{Symmetry breaking in the ground state}\label{sectsymm}

We now have the necessary tools to prove symmetry breaking. With
the results of Theorems \ref{largenthm} and \ref{decrthm} the
following is easily shown.

\begin{thm}[Symmetry breaking]\label{symbreak}
For all $0<\Omega< \Omega_c$ there is an $a_\Omega$ such that
$a\geq a_\Omega$ implies that {\em no} ground state of the functional
(\ref{gpfunct}) is an eigenfunction of the angular
momentum.
\end{thm}

\begin{proof}
{}Fix $0<\Omega< \Omega_c$. From Thm. \ref{largenthm} we see that
there exists an $N_\Omega$ independent of $a$ such that all vortex
states with $n\geq N_\Omega$ are unstable, and therefore cannot be
minimizers of $\E$. By the definition of the critical frequencies
(\ref{critdef}),
\begin{equation}
\min_{0\leq n<N_\Omega}\{E_n(a)-n\Omega\}>\min_{n\geq N_\Omega}
\{E_n(a)-n\Omega\}
\end{equation}
if
\begin{equation}
\Omega>\max_{0\leq j<N_\Omega}\frac
1{N_\Omega-j}\sum_{i=j}^{N_\Omega-1}\Omega_i(a).
\end{equation}
By Thm. \ref{decrthm} (and the remark after the proof) this can
always be fulfilled for $a$ large enough, so the ground state of
$\E$ cannot be an $n$-vortex.
\end{proof}

This shows that {\em no} minimizer of the GP functional is an
eigenfunction of the angular momentum. We can even show more,
namely that the absolute value of a minimizer is {\em not} a
radial function. To prove this, we need the following general
lemma.

\begin{lem}[Fourier series of $e^{ih(\varphi)}$]\label{fourier}
Let $h:[0,2\pi]\to \R$ be a measurable function. Then the set of
Fourier coefficients of $e^{ih}$ contains either only one or
infinitely many non-zero elements.
\end{lem}

\begin{proof}
Let $e^{ih(\varphi)}=\sum_n h_n e^{in\varphi}$, where the $h_n$'s
are the Fourier coefficients of $e^{ih}$. Let $n_0=\max\{n,
h_n\neq 0\}$ and $n_1=\min\{n,h_n\neq 0\}$, assuming that both are
finite. Since
\begin{equation}
1=\left| e^{i h(\varphi)}\right|^2=\sum_n k_n e^{in\varphi}\quad
{\rm with\ } k_n=\sum_m h_{n+m}\overline{h_m},
\end{equation}
we know that $k_n=\delta_{n0}$. But
$k_{n_0-n_1}=h_{n_0}\overline{h_{n_1}}\neq 0$, so $n_0=n_1$.
\end{proof}

\begin{cor}[Symmetry breaking, part 2]\label{symcor}
Let $0<\Omega<\Omega_c$ and $a\geq a_\Omega$, and let $\phi^{\rm
GP}$ be a minimizer of $\E$. Then $|\phi^{\rm GP}|$ is {\em not} a
radial function.
\end{cor}

\begin{proof}
Assume $|\phi^{\rm GP}|$ is radial. With $\widetilde
H=H_0+2a|\phi^{\rm GP}|^2-\mu$ we have $\widetilde H\phi^{\rm
GP}=0$. Because $\widetilde H$ commutes with $L$, it has
eigenfunctions $h_n(r)e^{in\varphi}$ with corresponding eigenvalue
$0$. Therefore \begin{equation} \phi^{\rm GP}(r,\varphi)=\sum_n
\lambda_n h_n(r)e^{in\varphi} \end{equation} for some
$0\neq\lambda_n\in\C$, and the sum is finite, since the ground
state energies of $\widetilde H$ restricted to subspaces of $L=n$
go to infinity as $n\to\infty$, and there can only be one
eigenfunction for each $n$. Choosing some interval $I\in\R_+$
where $|\phi^{\rm GP}|$ does not vanish, we can conclude with
Lemma \ref{fourier} that only one $h_n$ is unequal to zero in $I$.
But since the $h_n$'s do not vanish on some open set, this is true
on all of $\R_+$. Therefore $\phi^{\rm GP}$ has to be an
eigenfunction of $L$, contradicting Thm. \ref{symbreak}.
\end{proof}

Note that Thm. \ref{symbreak} implies in particular that the
minimizer of $\E$ is {\em not} unique (up to a constant phase),
and Corollary \ref{symcor} shows that even the absolute value is
not unique. By rotating a minimizer $\phi^{\rm GP}$ one obtains
again a minimizer, which is, at least except for exceptional
values of the rotation angle, different from the original one.

Numerical investigations \cite{BR99,CD99} indicate that the symmetry
breaking results from a splitting of an $n$-vortex into several
vortices with winding number $1$. I.e., one expects that $\phi^{\rm
GP}$ has $d$ distinct zeros of degree $1$, where $d={\rm
deg}\{\phi^{\rm GP}/|\phi^{\rm GP}|\}$ for large enough $r$. This
property was proved for large $a$ for the minimizer of models similar
to the GP functional \cite{BBH94,S01,AD01}. There is also experimental
evidence for vortex splitting \cite{MCWD00,ARVK01}.

\section{A density matrix functional}\label{sectdm}

We now introduce a new functional, which will be convenient in the
following. Firstly, to obtain a lower bound on the critical
parameter $a_\Omega$, and secondly, for studying a generalization
of the GP functional to a Bose gas with several components, which
will be done in the next section.

Analogously to the Gross-Pitaevskii functional we define the GP
{\it density matrix} (DM) functional as
\begin{equation}\label{defdm}
\Edm[\gamma]=\Tr [ H_0\gamma] + a \int \rho_\gamma(\x)^2 d^2\x.
\end{equation}
Here $\gamma$ is a one-particle density matrix, a positive
trace-class operator on $L^2(\R^2)$, and $\rho_\gamma$ denotes its
density. The ground state energy, the infimum of (\ref{defdm})
under the condition $\Tr[\gamma]=1$, will be denoted by
$\Endm(a,\Omega)$. It is clear that $\Endm\leq \Engp$. Using the
methods of \cite{LSY00} and \cite{BS01} one can prove the
following theorem.

\begin{thm}[Minimizer of $\Edm$]\label{thm5}
For each $0\leq \Omega<\Omega_c$ and $a\geq 0$ there exists a
minimizing density matrix for (\ref{defdm}) under the condition
$\Tr[\gamma]=1$. The density corresponding to the minimizer,
denoted by $\rdm$, is unique (and therefore a radial function),
and each minimizer also minimizes the linearized functional
\begin{equation}\label{deflin}
\Edm_{\rm lin}[\gamma]=\Tr [ (H_0+2a \rdm)\gamma] .
\end{equation}
\end{thm}

The uniqueness of the density results from the strict convexity of
$\Edm$ in $\rho_\gamma$. In general, the minimizing density matrix
need not be unique. However, for the functional (\ref{defdm}) we
can show that this is indeed the case.

\begin{thm}[Uniqueness of $\gdm$]\label{uniquedm}
The minimizer of $\Edm$, denoted by $\gdm$, is unique. Moreover,
it has finite rank.
\end{thm}

\begin{proof}
Being a minimizer of (\ref{deflin}), $\gdm$ can be
decomposed as
\begin{equation}
\gdm(\x,\x')=\sum_{j,k\geq 0} \lambda_{j,k} f_j(r) f_k(r') e^{i
(j\varphi-k\varphi')},
\end{equation}
where the  $f_k(r)e^{ik\varphi}$ are the ground states of
$H_0+2a\rdm$, and the sum is finite because of the discreteness of
the spectrum of this operator, implying finite rank of $\gdm$.
Moreover, there can be only one ground state for each angular
momentum, and $f_j(r)=r^j g_j(r)$ with $g_j(r)$ bounded and
strictly positive. Therefore
\begin{equation}
\rdm(r)=\sum_{j\geq 0} r^j \chi_j(r,\varphi),
\end{equation}
with
\begin{equation}
\chi_j(r,\varphi)=\sum_k \lambda_{j-k,k} g_{j-k}(r) g_k(r)
e^{i(j-2k)\varphi}.
\end{equation}
Hence each $\chi_j$ has to be independent of $\varphi$ as $r\to 0$, which
implies, together with Lemma \ref{fourier}, that $\lambda_{j,k}=0$ for
$j\neq k$. Moreover, $\lambda_{j,j}$ is determined by the unique density
$\rdm=\sum_{j\geq 0} r^{2j}\lambda_{j,j} g_j^2$, so $\gdm$ is unique.
\end{proof}

Analogously to minimizers of the GP functional, the DM density has
the following properties.

\begin{lem}[Properties of $\rdm$]
For $r>0$ we have $\rdm>0$ and $\rdm\in C^\infty$ if $V\in
C^\infty$. Moreover, $\|\rdm\|_\infty\leq \mdm/(2a)$, where $\mdm$
is the chemical potential of the DM theory, which is the ground
state energy of $\Edm_{\rm lin}$.
\end{lem}

\begin{proof}
Note that $\rdm=\sum_j \lambda_{j,j} f_j(r)^2$ with the notation
of the proof of Theorem \ref{uniquedm}, where $\lambda_{j,j}\geq
0$. The first two properties follow from this decomposition and a
bootstrap argument. Moreover, a direct computation gives
\begin{equation}
-\Delta\rdm\leq 2 \rdm(\mdm-2a\rdm - V(r)).
\end{equation}
Since $V(r)\geq 0$ this implies that $2a\rdm\leq \mdm$ by a
subharmonicity argument as in Lemma \ref{proplem}.
\end{proof}

An important consequence of the uniqueness of the minimizer of
$\Edm$ is the following corollary.

\begin{cor}[Non-equivalence of $\E$ and $\Edm$]
Assume that the minimizer of $\E$ is not unique, which is in
particular the case for $a\geq a_\Omega$. Then
$\Engp(a,\Omega)>\Endm(a,\Omega)$.
\end{cor}

\begin{proof}
This follows immediately from the uniqueness of $\gdm$.
\end{proof}

Note that in the case of non-uniqueness of $\phi^{\rm GP}$, the
rank of $\gdm$ is always greater or equal to two, and therefore
the ground state of $H_0+2a\rdm$ is degenerate. This holds in
particular in the whole region $a\geq a_\Omega$, not only for
isolated points or lines in the $(a,\Omega)$ plane. For the
non-rotating case, i.e. $\Omega=0$, $\Engp$ and $\Endm$ are equal
for all $a$. This remains true, if $\Omega$ is not too large.

\begin{prop}[Equivalence of $\E$ and $\Edm$ for small
$\Omega$]\label{equiv} Assume that
\begin{equation}
\Omega\leq \frac 14 \frac{\tilde\Omega^2}{C_{\tilde\Omega}+\mdm}
\end{equation}
for some $\Omega<\tilde\Omega<\Omega_c$. Then
$\Engp(a,\Omega)=\Endm(a,\Omega)$, and the minimizer of $\E$ has
zero angular momentum.
\end{prop}

Note that this Proposition implies a lower bound on $a_\Omega$.

\begin{proof}
Assume that $\psi$ is a ground state of $\Hdm\equiv H_0+2a\rdm$
with angular momentum $L\psi=m\psi$. Then
\begin{eqnarray}\nonumber
\mdm&=&\infspec\Hdm=\langle |\psi||-\Delta-\Omega m + \frac
{m^2}{r^2} + V(r)+2a\rdm||\psi|\rangle\\ &\geq&
\mdm+\langle\psi|\frac {m^2}{r^2}-\Omega m|\psi\rangle,
\end{eqnarray}
implying that
\begin{equation}
m\leq \frac \Omega{\langle\psi|r^{-2}|\psi\rangle}\leq \Omega
\langle\psi|r^2|\psi\rangle.
\end{equation}
Choosing some $\tilde\Omega$ with $\Omega<\tilde\Omega<\Omega_c$
we have $r^2\leq 4(C_{\tilde\Omega}+V(r))/\tilde\Omega^2$ by
(\ref{propv}), and hence
\begin{eqnarray}\nonumber
\left(1-\frac{\Omega^2}{\tilde\Omega^2}\right)
\langle\psi|r^2|\psi\rangle &\leq& \frac 4{\tilde\Omega^2}\left(
C_{\tilde\Omega}+\langle\psi|V(r)-\frac{\Omega^2}{4}r^2|\psi\rangle\right)\\
&\leq& \frac
4{\tilde\Omega^2}\left(C_{\tilde\Omega}+\mdm-\Omega\right),
\end{eqnarray}
where we have used that the ground state energy of $-\Delta-\Omega
L+ \Omega^2 r^2/4$ is $\Omega$. Therefore
\begin{equation}
\Omega \langle\psi|r^2|\psi\rangle\leq \frac
{4\Omega\left(C_{\tilde\Omega}+\mdm-\Omega\right)}{\tilde\Omega^2-\Omega^2}<1
\end{equation}
if
\begin{equation}
\Omega\leq \frac 14 \frac{\tilde\Omega^2}{C_{\tilde\Omega}+\mdm},
\end{equation}
showing that any ground state of $\Hdm$ necessarily has angular
momentum $m=0$.
\end{proof}

\section{The multi-component Bose gas}\label{sectmult}

We now consider the Gross-Pitaevskii theory of a rotating Bose gas
with $n_c$ different components, or equivalently, a Bose gas
consisting of particles with spin $(n_c-1)/2$. The natural
generalization of the Gross-Pitaevskii functional is
\begin{equation}
\E_\nc[\phi_1,\dots,\phi_\nc]=\sum_{i=1}^\nc
\langle\phi_i|H_0\phi_i\rangle+a \sum_{1\leq i,j\leq
\nc}\int|\phi_i|^2|\phi_j|^2,
\end{equation}
which has to be minimized under the constraint $\sum_{i=1}^\nc
\int|\phi_i|^2=1$. The corresponding ground state energy will be
denoted by $\Engp_\nc(a,\Omega)$.

Using standard methods, one can show that for all values of
$\nc\in\N$, $0\leq \Omega<\Omega_c$ and $a\geq 0$ there exist
minimizing functions $\phi_1^{\rm GP}$,\dots,$\phi_\nc^{\rm GP}$
for $\E_\nc$. The proof goes analogously to the proof of Thm.
\ref{thm5}, noting that $\E_\nc$ can be considered as the
restriction of $\Edm$ to density matrices of rank less than or
equal to $n_c$. However, since this set is not convex, we can in
general not conclude that the density of a minimizer,
$\sum_{i=1}^\nc |\phi^{\rm GP}_i|^2$, is unique, as it was the
case for the DM functional.

We see that for all values of $a$, $\Omega$ and $\nc$ we always
have $\Endm\leq \Engp_\nc\leq \Engp$. Denoting
\begin{equation}
\ndm(a,\Omega)={\rm rank\, }\gdm,
\end{equation}
which we showed to be finite in Thm. \ref{uniquedm}, we can
distinguish the following cases.

\begin{thm}[Minimizers of the multi-component GP functional]
Let $\phi^{\rm GP}_1$,\dots, $\phi^{\rm GP}_\nc$ be minimizers of
$\E_\nc$.
\begin{itemize}
\item[(i)] If $\nc\geq\ndm$, then
$\Endm(a,\Omega)=\Engp_\nc(a,\Omega)$. Moreover,
\begin{equation}\label{gdmnc}
\sum_{i=1}^\nc |\phi^{\rm GP}_i\rangle\langle \phi^{\rm GP}_i|
=\gdm,
\end{equation}
implying that the $\phi^{\rm GP}_i$'s can be written as $\phi^{\rm
GP}_i=\sum_{j=1}^\ndm A_{ij}\psi_j$, where $\gdm=\sum_{i=1}^\ndm
|\psi_i\rangle\langle \psi_i|$, $\psi_i$ orthogonal, and $A$ is an
$\nc\times \ndm$-matrix with $A^\dagger A=1$.
\item[(ii)] If $\nc<\ndm$, then
$\Endm(a,\Omega)<\Engp_\nc(a,\Omega)$.
\item[(iii)]
If $\nc\geq 2$, $a\geq a_\Omega$, then
$\Engp_\nc(a,\Omega)<\Engp(a,\Omega)$, and the minimizers
$\phi^{\rm GP}_i$ are {\em not} all equal; i.e., $\sum_{i=1}^\nc
|\phi^{\rm GP}_i\rangle\langle \phi^{\rm GP}_i|$ has at least rank
$2$.
\end{itemize}
\end{thm}

\begin{proof}
As remarked earlier, $\Endm\leq \Engp_\nc$. To prove (i), we write
$\gdm=\sum_{i=1}^\ndm  |\psi_i\rangle\langle\psi_i|$, where the
$\psi_i$ are orthogonal (but not necessarily normalized). Using
$\phi_i=\psi_i$ for $i\leq \ndm$ and $\phi_i=0$ for $\ndm<i\leq
\nc$ as trial functions, we obtain $\Endm\geq\Engp_\nc$. Since
$\gdm$ is unique, (\ref{gdmnc}) follows. (ii) is a trivial
consequence of the uniqueness of $\gdm$.

For $a\geq a_\Omega$, we know from Corollary \ref{symcor} that
there are at least two different minimizers, $\phi^{(1)}$ and
$\phi^{(2)}$, for $\E$, whose absolute values are not the same.
For $\nc\geq 2$ we use as trial functions $\phi_1=\phi^{(1)}$,
$\phi_i=\phi^{(2)}$ for $i\geq 2$ to obtain $\Engp_\nc<\Engp$.
Therefore the minimizers for $\E_\nc$ cannot be all equal, and
(iii) is proved.
\end{proof}

An important consequence of part (iii) of the theorem above is
that the GP ground state energy depends non-trivially on the
number of spin-components, at least in the symmetry breaking
regime $a\geq a_\Omega$. Moreover, there is a clear separation
between different spin-components, their individual densities
$|\phi_i^{\rm GP}|^2$ can never be all equal.

\section{The special case $V(r)=r^2$}\label{sectharm}

In the case of a harmonic potential $V(r)=r^2$ the theorems above
are more explicit, setting $\Omega_c=2$ and $C_\Omega=0$.
Moreover, the special case $\Omega=\Omega_c=2$ is easy to treat:
only for $a=0$ there is a minimizer for $\E$ (in fact there are
infinitely many), whereas for $a>0$ there is no minimizer, and
also all $n$-vortices are unstable.

The following bound on the energies $E_n(a)$ can be easily
obtained, and will be used several times in the considerations
below.

\begin{lem}[Upper bound on $E_n(a)$]\label{uplem}
\begin{equation}\label{uppen}
E_n(a)\leq 2(n+1)\sqrt{1+\frac a {b_n (n+1)}}
\end{equation}
with $b_n=2\pi 4^n (n!)^2/(2n)!$.
\end{lem}

\begin{proof}
This follows from a trial function of the form $C r^n \exp(-c
r^2)$, where $C$ is a normalization constant and $c$ is to be
optimized.
\end{proof}

The estimate above implies that, for $\tilde\mu$ the chemical
potential corresponding to the minimizer of $\En$,
\begin{equation}\label{mutwon}
\tilde\mu-2n\leq 2(E_n-2n)-2\leq
2\left(1+2\sqrt{a\frac{(n+1)}{b_n}}\right).
\end{equation}
This will be useful for an upper bound on $f$, since in the case
of $V(r)=r^2$ we can improve the estimate (\ref{fsup}) by
\begin{equation}\label{fimpr}
\|f\|_\infty^2\leq \frac 1{2a}(\tilde\mu-2n).
\end{equation}
The proof is analogous to the one of (\ref{fsup}), using in
addition that $n^2/r^2+r^2\geq 2n$.

We now consider the stability of $n$-vortices $\phi$, i.e.,
solutions to (\ref{station}) of the form (\ref{vortex}). In
addition to the results of Section \ref{sectvort} we can state
another proposition.

\begin{prop}[Instability for small $a$]
Assume that $a< \pi n (2-\Omega)$. Then $\phi$ is unstable.
\end{prop}

\begin{proof}
Let $n\geq 1$, and let $w(\x)=\sqrt{1/\pi} \exp(-r^2/2)$ be the
ground state of $H_0$. Then
\begin{equation}
Q(w)=2-\mu+4 a \int w^2 |\phi|^2\leq n(\Omega-2)+\frac a\pi,
\end{equation}
where we used Cauchy-Schwarz and the fact that $\mu\geq
2-n(\Omega-2)+2 a \int |\phi|^4$.
\end{proof}

Moreover, we can improve Theorem \ref{largenthm} in the following
way.

\begin{thm}[Instability for large $n$, harmonic potential]\label{largen2}
Assume that $n\geq 10$ and
\begin{equation}\label{largen}
\tilde\mu\geq \frac{n\Omega}{1-d_n},
\end{equation}
where $d_n$ is a monotone decreasing function of $n$, with $d_n<1$
for $n\geq 10$, namely
\begin{equation}\label{defdn}
d_n=\min\left\{\frac 2{e^2}+\frac{2\pi\Gamma(n+\half)^2}{n^2
\Gamma(n)^2}+2^{1-n}\left(n!-\Gamma(n+1,2)\right),\frac{19}{n}\right\}.
\end{equation}
Then $\phi$ is unstable.
\end{thm}

Note that since $\tilde\mu>2n$, (\ref{largen}) is in particular
fulfilled if $d_n\leq 1-\Omega/2$.

\begin{proof}
Let $n\geq 1$, and let $w_1\in H^1(\R^2)$ be radial and
normalized. Let $T=\langle w_1|-\Delta w_1\rangle$ and $X=\langle
w_1|r^2 w_1\rangle$, and define $w$ by
\begin{equation}
w(\x)=c_n \tilde\mu^{1/2} w_1(c_n\tilde\mu^{1/2} r),
\end{equation}
with $c_n$ given in (\ref{cn}). Using (\ref{fimpr}) and
(\ref{gbound}), we can estimate
\begin{eqnarray}\nonumber
Q(w)&\leq& n\Omega+\tilde\mu\left(c_n^2 T+\frac
1{c_n^2\tilde\mu^2}X-1\right)\\ && +2(\tilde\mu-2n)\int |w(\x)|^2
\min\{1,(r^2\tilde\mu c_n^2)^n\} d^2\x.
\end{eqnarray}
With $N=\int_{r\geq 1} |w_1|^2$ and $M=\int_{r\leq
1}|w_1|^2r^{2n}$ this gives, using $\tilde\mu\geq 2(n+1)$ in front
of $X$,
\begin{equation}\label{79}
Q(w)\leq n\Omega+\tilde\mu\left(c_n^2
T-1+2\left(N+M\right)\right)+\left(\frac 1{c_n^2 2(n+1)}X-4n
(N+M)\right).
\end{equation}
Now if we choose $w_1(r)=(2/\pi)^{1/2}\exp(-r^2)$ the last term in
(\ref{79}) is negative, and the computation of $T$, $N$ and $M$
yields
\begin{equation}
Q(w)< n\Omega-\tilde\mu(1-d_n),
\end{equation}
where $d_n$ is the first part in the parenthesis in (\ref{defdn}).
For large $n$, this can be improved by choosing
$w_1(r)=(35/9\pi)^{1/2}[1-r^{3/2}]_+$, which gives
\begin{equation}
Q(w)< n\Omega-\tilde\mu\left(1-\frac{19}n\right).
\end{equation}
\end{proof}

In \cite{CD99} the authors used a similar method to the one of
Thm. \ref{largen2} and a particular assumption on the form of the
vortex state $\phi$ to obtain a $d_n$ in (\ref{largen}) that is
less than $1$ if $n\geq 2$.

We conjecture that in the case of an harmonic potential an
$n$-vortex with $n\geq 2$ is unstable, for all values of
$\Omega\geq 0$ and $a> 0$ . However, we can prove this only for
$\Omega\leq 1$. Namely, if we insert $V(r)=r^2$ in (\ref{227}),
set $d=2$, use the improved bound (\ref{fimpr}) and $\mu\geq
2-n(\Omega-2)+2 a \int |\phi|^4$ this shows the negativity of
$Q(w)$ as long as $n\geq 2$ and
\begin{equation}\label{condomega}
\Omega<1+\frac 1{2n}\left(1+a \int |\phi|^4\right)
\end{equation}
is satisfied. This implies that all vortices with $n\geq 2$ and
$a\geq 0$ are unstable as long as $\Omega\leq 1$.

As a consequence of the considerations above we can state an
explicit condition on $a$ where an $n$-vortex is necessarily
unstable, using the general lower bound
\begin{equation}
\int |\phi(\x)|^4 d^2\x \geq \frac 4{9\pi}\frac {\left(\int
|\phi(\x)|^2
 d^2\x\right)^3}{\int |\phi(\x)|^2 |\x|^2 d^2\x},
\end{equation}
which can easily be proved using elementary calculus of
variations. Moreover, since in two dimensions $\int |\phi|^4$
scales as $1/({\rm length})^2$, the virial theorem implies for the
minimizer $f$ of $\En$
\begin{equation}
\int f(r)^2 r^2 d^2\x=\frac 12 E_n(a).
\end{equation}
Hence we only need an upper bound on $E_n(a)$, which is given in
Lemma \ref{uplem}, to obtain a condition on $a$ for validity of
(\ref{condomega}).

The critical frequencies $\Omega_n(a)$, defined in
(\ref{critdef}), have the following properties.

\begin{lem}[Properties of critical frequencies]\label{propom}
For all $n\in \N_0$ we have $\Omega_n(0)=2$ and for all $a\geq 0$
$\lim_{n\to\infty} \Omega_n(a)=2$. Moreover,
\begin{equation}\label{omprime}
\Omega'_n(0)=-\frac 1 {4^{n+1}
\pi} \frac {(2n)!}{n!(n+1)!}<\Omega'_{n+1}(0)<0.
\end{equation}
\end{lem}

\begin{proof}
The first assertion follows from $E_n(0)=2(n+1)$. Using the
harmonic oscillator eigenstates $\chi_n(r)=\sqrt{1/\pi n!}r^n
\exp(-r^2/2)$ as trial functions the second assertion is proved by
\begin{equation}
\Omega_n\leq 2+a\int |\chi_{n+1}|^4\qquad {\rm and}\quad
\Omega_n\geq 2 - a\int |\chi_n|^4,
\end{equation}
noting that
$\int|\chi_n|^4=O(n^{-1/2})$ as $n\to\infty$. To prove
(\ref{omprime}) we use the Feynman-Hellmann principle to calculate
\begin{equation}
\Omega_n'(a)=\int |f_{n+1}|^4 - \int |f_n|^4,
\end{equation}
where
$f_n$ is the minimizer of $\En$. For $a=0$ we have $f_n=\chi_n$,
yielding (\ref{omprime}).
\end{proof}

In the special case of a harmonic potential, the results of Thm.
\ref{decrthm} can be improved. We get the following bounds on the
critical frequencies.

\begin{thm}[Bounds on $\Omega_n$, harmonic potential]\label{decrthm2}
For all $n\in\N_0$
\begin{eqnarray}\label{crit2}
&&\Omega_n(a)\leq (2n+1)\frac{2\pi e}a
\left(1+\sqrt{\frac{2a}{\pi}}\right) \left(3+\left[\ln\left(\frac
a{2\pi e^2}\right) \right]_+\right),\\ \label{lowom} &&
\Omega_n(a)\geq \frac{2n+1}{(n+2)\sqrt{1+\frac a {b_{n+1}
(n+2)}}},
\end{eqnarray}
with $b_n$ given in Lemma \ref{uplem}.
\end{thm}

\begin{proof}
We proceed as in Thm. \ref{decrthm}, but now use the improved
estimate (\ref{fimpr}) to replace (\ref{312}) by
\begin{equation}
\frac{\partial E_n}{\partial n}\leq \frac{\pi e}a (\tilde\mu-2n)
\max\left\{2n+1,n\ln\left(\frac{c_n^2 a}{2 \pi}\right)\right\}.
\end{equation}
Inserting (\ref{mutwon}) and using $(n+1)/b_n\leq (2\pi)^{-1}$ for
$0\leq n\leq 1$ and $c_n^{2n}\leq e$ this gives for $\Omega_0$
\begin{equation}
\Omega_0(a)\leq \frac{2\pi e}a
\left(1+\sqrt{\frac{2a}{\pi}}\right) \max\left\{3,1+\ln\left(\frac
a{2\pi}\right) \right\},
\end{equation}
and using Lemma \ref{omegan0} we obtain (\ref{crit2}).

For the lower bound we proceed as in (\ref{314}) and note that
$\int f_{n+1}^2 r^2=E_{n+1}(a)/2$ by the virial theorem.
(\ref{lowom}) is obtained by inserting the bound (\ref{uppen}) for
$E_{n+1}(a)$.
\end{proof}

Numerical investigations in \cite{GP99} indicate that for all
$n\geq 0$, $\Omega_n$ is strictly monotone decreasing in $a$, and,
for $a>0$, $\Omega_n<\Omega_{n+1}$, i.e., $E_n$ is convex in $n$.
Note that the theorem above states that $\Omega_n$ behaves at most
as $a^{-1/2}\ln a$ for large $a$, in accordance with previous
considerations (see \cite{CD99} and references therein).

We can now use the results above to derive explicit upper and
lower bounds on $a_\Omega$. Denote
\begin{equation}\label{defxi}
\Xi(a)=\frac{2\pi e}a \left(1+\sqrt{\frac{2a}{\pi}}\right)
\left(3+\left[\ln\left(\frac a{2\pi e^2}\right) \right]_+\right),
\end{equation}
which is a strictly monotone decreasing function of $a$. We know
from Thm. \ref{largen2} and (\ref{condomega}) that all
$n$-vortices are unstable for $n\geq N_\Omega$, where
\begin{equation}\label{723}
N_\Omega=\left\{\begin{array}{cl} 2 & \mbox{for $\Omega\leq 1$} \\
\frac {38}{2-\Omega} & \mbox{for $\Omega > 1$.} \\
\end{array}\right.
\end{equation}
Using the bound on the critical frequencies (\ref{crit2}), we see
by analogous considerations as in the proof of Thm. \ref{symbreak}
that symmetry breaking occurs if
\begin{equation}
\Xi(a)\leq \frac\Omega{2N_\Omega-1}.
\end{equation}

From Thm. \ref{equiv} we see that
$\Endm(a,\Omega)=\Engp(a,\Omega)$ if $\Omega\leq 1/\mdm$.
Moreover, it is easy to see that the same holds for $\Omega\leq
2-a/\pi$. Namely, using Cauchy-Schwarz and the fact that
$\int(\rdm)^2$ is monotone decreasing in $a$ (because of concavity
of $\Endm$ in $a$) we have, for $\Hdm=H_0+2a \rdm$,
\begin{equation}
\infspec\Hdm\restriction_{L=0}\leq 2+2a\int\rdm\chi^2 <
2+2a\int\chi^4 = 2+\frac a\pi,
\end{equation}
where $\chi(\x)=\sqrt{1/\pi}\exp(-r^2/2)$ is the ground state of
$H_0$. Moreover,
\begin{equation}
\Hdm\restriction_{|L|\geq 1}> 4-\Omega,
\end{equation}
showing that, for $a\leq \pi(2-\Omega)$, $\Hdm$ has a unique
ground state with zero angular momentum which necessarily also
minimizes the GP functional.

Since the minimizer of the GP functional is unique and therefore
an angular momentum eigenfunction as long as $\Endm=\Engp$, we
obtain as a consequence a lower bound on $a_\Omega$, using
$\mdm\leq 2\Endm(a,\Omega)\leq 2 E_0(a)$ and the bound on $E_0(a)$
given in (\ref{uppen}). Thus we have proved the following Theorem:

\begin{thm}[Bounds on $a_\Omega$]
In the case of a harmonic potential, the critical parameter for
symmetry breaking fulfills the bounds
\begin{equation}
a_\Omega\leq \Xi^{-1}\left(\frac\Omega{2N_\Omega-1}\right),
\end{equation}
with $\Xi$ defined in (\ref{defxi}) and $N_\Omega$ given in
(\ref{723}), and
\begin{equation}
a_\Omega\geq \pi\, \max\left\{2-\Omega,\frac
1{8\Omega^2}-2\right\}.
\end{equation}
\end{thm}

\section*{Acknowledgments}
The author would like to thank Elliott Lieb and Jakob Yngvason for
fruitful discussions and helpful comments. Financial support by
the Austrian Science Fund in the form of an Erwin
Schr\"odinger Fellowship is gratefully acknowledged.

\end{document}